\documentclass[prl,superscriptaddress,twocolumn,preprintnumbers,a4,showpacs,floatfix]{revtex4}
\usepackage{amsmath, amssymb, psfrag, tabls, dcolumn, bm, color}
\usepackage[sort&compress]{natbib}
\usepackage[pdftex]{graphicx}
\usepackage{graphicx}
\usepackage{bm}

\newcommand\smallurl[1]{{\tiny \url{#1}}}

\newcommand{\be}{\begin{equation}}
\newcommand{\ee}{\end{equation}}
\newcommand{\bea}{\begin{equation*}}
\newcommand{\eea}{\end{equation*}}
\newcommand{\ba}{\begin{array}}
\newcommand{\ea}{\end{array}}
\newcommand{\beqa}{\begin{eqnarray}}
\newcommand{\eeqa}{\end{eqnarray}}
\newcommand{\beqaa}{\begin{eqnarray*}}
\newcommand{\eeqaa}{\end{eqnarray*}}

\newcommand{\matr}{\left( \begin{array}}
\newcommand{\ematr}{\end{array} \right)}

\newcommand{\rb}{\mbox{\boldmath $r$}}

\newcommand{\nb}{\mbox{\boldmath $n$}}

\newcommand{\Fb}{\mbox{\boldmath $F$}}

\newcommand{\Tb}{\mbox{\boldmath $T$}}

\newcommand{\der}{{\rm d}}

\newcommand{\lsim}{{\;\raise0.3ex\hbox{$<$\kern-0.75em\raise-1.1ex\hbox{$\sim$}}
\;}}
\newcommand{\gsim}{{\;\raise0.3ex\hbox{$>$\kern-0.75em\raise-1.1ex\hbox{$\sim$}}
\;}}

\begin{document}

\title{Efficient approach for simulating distorted materials}

\author{Pekka Koskinen\footnote{Corresponding author}}
\email[email:]{pekka.koskinen@iki.fi}
\address{NanoScience Center, Department of Physics, University of Jyv\"askyl\"a, 40014 Jyv\"askyl\"a, Finland}

\author{Oleg O. Kit}
\address{NanoScience Center, Department of Physics, University of Jyv\"askyl\"a, 40014 Jyv\"askyl\"a, Finland}


\begin{abstract}
The operation principles of nanoscale devices are based upon both electronic and mechanical properties of materials. Because these properties can be coupled, they need to be investigated simultaneously. At this moment, however, the electronic structure calculations with custom-made long-range mechanical distortions are impossible, or expensive at best. Here we present a unified formalism to solve exactly the electronic structures of nanomaterials with versatile distortions. We illustrate the formalism by investigating twisted armchair graphene nanoribbons with the least possible number of atoms. Apart from enabling versatile material distortions, the formalism is capable of reducing computational costs orders of magnitude in various areas of science and engineering.
\end{abstract}

\maketitle

%
%
Bloch's theorem has been the propulsive force of computational materials research for more than $80$ years\cite{bloch_ZP_28}, today as important as ever. While the theorem still associates with Bravais lattices and translational symmetry, nanoscience has brought us low-dimensional structures, tubes, tori, wires, and membranes, which get twisted, bent, wrapped, and rippled in experiments. Translational symmetry hides deceiving simulation constraints, since materials cannot distort the way they would prefer, and restricts realistic modeling of nanoelectromechanical components.

Distortions are relevant in a number of topical material systems: polymers, double helices like DNA, lipid bilayers, nanoscrolls, nanocoils, nanowires, and, especially, carbon nanostructures including fullerenes, carbon nanotubes (CNTs), graphene, and graphene nanoribbons (GNRs), to mention a few.\cite{castro_neto_MT_10} For example, materials with high aspect ratio like CNTs and GNRs get bent\cite{shenoy_PRL_08,malola_PRB_08b,bets_NR_09} and thin sheets like graphene get rippled \cite{meyer_nature_07,bao_nnano_09,shenoy_PRL_08}, unless carefully placed on a support. Classical modeling of distortions is a mature subject \cite{landau_lifshitz,bhuang_PRL_09,kudin_PRB_01}, but while classical interaction potentials and finite element methods give materials' mechanical properties, they are useless for electronic properties. In nanoscience quantum-mechanical modeling is preferred.

How can we include quantum mechanics into these distortion simulations? For decades chemists have used group theory and molecular symmetries to reduce computational costs. In computational materials physics, symmetries beyond translation have been used mainly for chiral carbon nanotubes, in work pioneered by White, Robertson and Mintmire\cite{white_PRB_93}, followed by Popov\cite{popov_NJP_04} and Dumitric{\u a}\cite{zhang_APL_08}, with co-workers. Nanotubes are natural because chiral symmetry itself suggests ``symmetry adaption''; it is, however, less evident to \emph{break} the symmetry and investigate the elastic properties in a broader sense.

In this Letter we shall present a compact, exact, and flexible formalism to solve the electronic structure of nanomaterials with custom-made distortions. By expanding the concepts of periodicity and simulation cells, the formalism can also reduce computational costs, even in classical materials modeling.

\def\trans{\hat{\mathcal{T}}}
\def\transt{\hat{\mathcal{T}}_{\Tb}}
\def\sop{\mathcal{S}}
\def\sopn{\mathcal{S}^{\bm n}}
\def\dsopn{\hat{D}(\mathcal{S}^{\bm n})}

The formalism is obtained by revising Bloch's theorem, and the derivation is straightforward. Consider electrons in a potential $V(\rb)$ that remains invariant in symmetry operations $\sopn$,
\begin{equation}
\dsopn V(\rb)=V(\mathcal{S}^{-{\bm n}}\rb)=V(\rb).
\end{equation}
The operation $\sopn$, with inverse $\mathcal{S}^{-{\bm n}}$, is a succession of $n_i$ times operation $\sop_i$ for all $i=1,2,\ldots$, that is $\sopn=\sop_1^{n_1} \sop_2^{n_2} \cdots$ with ${\bm n}=(n_1,n_2,\cdots)$. $\sop_i$ can be \emph{any} symmetry operation, such as translation, rotation, reflection, inversion, joined translation+rotation, or joined translation+reflection, to mention six, and they should form an abelian group.

Let us give a couple of familiar examples. With bulk $\sop_i$'s are three translations; for a benzene ring (C$_6$H$_6$) $\sop_1$ could be a two-, three-, or six-fold rotation around the symmetry axis; for an achiral carbon nanotube $\sop_1$ could be a translation along the symmetry axis and $\sop_2$ could be, say, an $M$-fold rotation around the symmetry axis; for polyethene ([--CH$_2$CH$_2$--]$_n$) $\sop_1$ could be a  translation across one CH$_2$ unit followed by a reflection (normally $\sop_1$ would be a translation across the whole CH$_2$CH$_2$ unit).

Now, returning to the derivation, since the transformations are isometric, the kinetic energy term in the Hamiltonian
\begin{equation}
\hat{H}=-\frac{\hbar^2}{2m_e}\nabla^2 + V(\rb)
\end{equation}
remains invariant, and $\hat{H}$ commutes with $\dsopn$, the two operators consequently sharing the same eigenstates. We denote these eigenstates $\psi_{a {\bm \kappa}}(\rb)$, with ${\bm \kappa}=(\kappa_1,\kappa_2,\ldots)$. Hence we have
\begin{equation}
\hat{D}(\sop_1^{p}) \psi_{a {\bm \kappa}}(\rb) = \lambda(\kappa_1)^p\psi_{a {\bm \kappa}}(\rb), \;(p \text{ integer})
\end{equation}
where $\lambda(\kappa_1)$ is the eigenvalue of $\hat{D}(\mathcal{S}_1)$. Since electron density remains invariant under symmetry operations,
\begin{equation}
|\lambda(\kappa_1)^p\psi_{a {\bm \kappa}}(\rb)|=|\psi_{a {\bm \kappa}}(\rb)|,
\end{equation}
we get $\lambda(\kappa_1)=\exp[i\alpha(\kappa_1)]$. Now we impose periodic boundary conditions by making the group cyclic $\sop_1^{M_1} \equiv \hat{1}$, and get $\alpha(\kappa_1) \equiv \kappa_1=2\pi m_1/M_1$, with integers $M_1$ and $m_1 \in [0,M_1-1]$.

\renewcommand{\arraystretch}{1.3}
\begin{table*}
\caption{Selected examples on the approach usage. The coordinates in parentheses mean the sense of the symmetry operation.}
\label{tab:usage}
\begin{tabular}{p{5.5cm}p{0.5cm}p{11.5cm}}
\hline & \\[-11pt]\hline
\hspace{1.5cm}Operations & & \hspace{2cm} Examples of usage \& notes \\
\hline
$\sop_1(z)$ rotation & &
bend tubes, wires, ribbons \cite{malola_PRB_08b} \\

$\sop_1(z)$ rotation and $\sop_2(z)$ translation &  &
bend membranes, slabs \\

$\sop_1$ joined rotation$(z)$ + translation$(z)$ &  &
twist nanotubes, wires, ribbons, DNA, simulate springs and coils \cite{lawler_PRB_06,pan_JAP_02,ko_APL_04} \\

$\sop_1(x)$ and $\sop_2(y)$ two rotations around the same origin & &
simulate spherical symmetry; solid and liquid membranes, such as mono- and multilayer graphene and lipid bilayers. ($\sop_1(x)$ and $\sop_2(y)$ commute approximately if rotation angles are small, and curvature can be treated as a \emph{local} property.)\\

$\sop_1(x)$, $\sop_2(y)$ rotations around different origins & &
simulating arbitrary Gaussian curvature, like saddle structures (approximate treatment, like above) \\

$\sop_1$ translation$(x)$+reflection ($yz$) & &
structures with repeating units $\ldots ABBA\ldots$, using $AB$ unit; many waves can be simulated using half the wavelength\cite{ABBA-note} \\

$\mathcal{S}_1(x)$, $\mathcal{S}_2(y)$ translations,
$\mathcal{S}_3(xy)$ reflection [+optional translation $(x$ or $y)$]& &
computational surface science; reflection doubles the surface slab thickness with half the number of atoms\\

normal point group symmetries & &
finite symmetric molecules and clusters \\

\hline & \\[-11pt]\hline
\end{tabular}
\end{table*}

By repeating the above steps for the remaining symmetry operations, we obtain a revised Bloch's theorem: in a potential, that is invariant in symmetry operations $\sopn$, the energy eigenstates $\psi_{a {\bm \kappa}}$ at $\rb$ and at $\rb'=\mathcal{S}^{-\bm n} \rb$ differ by a phase factor $\exp(i {\bm \kappa}\cdot \nb)$,
\begin{equation}
\hat{D}(\mathcal{S}^{\bm n})\psi_{a {\bm \kappa}}(\rb)=\psi_{a {\bm \kappa}}(\mathcal{S}^{-{\bm n}} \rb)=\exp(i{\bm \kappa}\cdot \nb)\psi_{a {\bm \kappa}}(\rb).
\label{eq:bloch}
\end{equation}
This implies that wave functions only in one unit cell---whatever its shape---determine the electronic structure of the whole, extended system.

You may recognize that the theorem above is nothing but Bloch's theorem, merely written with unconventional symbols. Indeed, many things remain as usual. Energy eigenstates $\psi_{a {\bm \kappa}}$ automatically fulfill Eq.(\ref{eq:bloch}), once written in a revised version of Bloch basis,
\begin{equation}
|{\bm \kappa},\mu \rangle \equiv \varphi_{\mu}({\bm \kappa},\rb)=\frac{1}{\sqrt{N}}\sum_{\bm n} \exp(-i{\bm \kappa}\cdot \nb)\hat{D}(\mathcal{S}^{\bm n}) \varphi_\mu(\rb),
\label{eq:bloch-basis}
\end{equation}
where $\varphi_\mu(\rb)$ are local orbitals and $\sum_{\bm n}1=N$ is the number of unit cells. The Bloch basis gives Hamiltonian diagonal in ${\bm \kappa}$,
\begin{equation}
\langle {\bm \kappa},\mu|\hat{H}|{\bm \kappa'},\nu\rangle=\delta({\bm \kappa}-{\bm \kappa'})\sum_{\bm n} \exp(-i{\bm \kappa}\cdot \nb)H_{\mu\nu}({\bm n})
\end{equation}
with
\begin{equation}
H_{\mu\nu}({\bm n})=\int \varphi_\mu^*(\rb)\hat{H} \left[ \hat{D}(\mathcal{S}^{\bm n})\varphi_\nu(\rb) \right] \der^3 r,
\label{eq:h-mel}
\end{equation}
and similarly for overlap matrix elements. The total energy expressions remain the same, we only use a set of ${\bm \kappa}$-points instead of ${\bm k}$-points (extra symmetries reduce the set).\cite{martin_book} Because forces are calculated as parametric derivatives of the total energy, molecular dynamics works normally and energy is conserved; simulation cell dynamics, however, are different, and the concept of pressure needs redefinition. Finally, the theorem works with any electronic structure method, whether it is \emph{ab initio} or not, whether it uses real-space grids or local orbitals (plane waves are tricky), or whether the approach is numerical or analytical.

Some things, however, do change in the revised Bloch's theorem. For bulk the periodic boundary condition is an approximation, whereas here some symmetries may form cyclic groups in reality, as in benzene. For cyclic groups the ${\bm \kappa}$-point sampling is more restricted; in the above example of an achiral carbon nanotube the translational component $\kappa_1$ can be freely sampled between $[-\pi,\pi]$ (because for $\mathcal{S}_1$ periodicity is an approximation), but the rotational component accepts only the discrete values $\kappa_2=2\pi m/M$, $m=0,1,\ldots,M-1$. Group multiplication tables for $\mathcal{S}_i$ that have identities like $\mathcal{S}_i^{l_i}\!\!=\!\mathcal{S}_j^{l_j}$ make the sampling of the components of ${\bm \kappa}$ coupled, for then we must have $l_i \kappa_i = l_j \kappa_j + 2 \pi m$ ($l_i,m$ integers). The connection between ${\bm \kappa}$- and ${\bm k}$-points is as follows. If $\mathcal{S}_1$ is a translation along ${\bm L}^1$, then $\mathcal{S}_1 \psi_{a{\bm \kappa}}$  should give the same phase as $\mathcal{T}({\bm L}^1) \psi_{a{\bm k}}$, if $\psi_{a{\bm \kappa}}$ and $\psi_{a{\bm k}}$ are the same physical states. Hence $\exp({-i\kappa_1})=\exp({-i{\bm k}\cdot {\bm L}^1})$, and, in general, $\kappa_i = \sum_{j=1}^{3} L^i_j k_j$ for $i=1,2,3$.
%

The formalism gives surprises, too. An atom can perform work on itself. This is because the total force on an atom exerted by its own periodic images---if rotations are involved---may differ from zero. Furthermore, a force $\Fb_{JI}$ on atom $I$ exerted by atom $J$ may not be the counterforce to the force on atom $J$ exerted by atom $I$, that is $\Fb_{JI}\neq - \Fb_{IJ}$; Newton's third law appears invalid. These unorthodoxies are \emph{not} bugs; remember that we simulate the whole extended system, and an atom $I$ in the primitive unit cell is different from the atom $I$ in a different unit cell---an artifact of atom indexing. Finally, note that if $\mathcal{S}_i$ contains rotations, also local orbitals rotate; this is implicit in the operation $\hat{D}(\mathcal{S}_i)\varphi_\mu(\rb)$ in Eq.(\ref{eq:bloch-basis}).



We implemented this formalism using local basis in the density-functional tight-binding software \texttt{hotbit}\cite{koskinen_CMS_09,hotbit_wiki}, and tested it with many finite and extended structures. We omit the details of the implementation here, and just comment on three things. First, the standard methods of electrostatics, like Ewald summation, are invalid since flexibility is required; we chose to use multipoles as they easily lend themselves for rotations and reflections. Second, the implementation can be done so that only the mappings
\begin{equation}
\rb' = \mathcal{S}_i \rb
\end{equation}
are needed to build new symmetries; this requires just a couple of lines new code. Implementation generally is not hard, but it may be nontrivial for codes already build upon translational symmetry. Third, implementation has a negligible computational overhead compared to translational symmetry (see Table~\ref{tab:compare}). Certain manipulations take more time, but the most CPU-intensive parts remain as usual.

So far our discussion has been abstract, but what can we do with the formalism in practice? While it may seem that we require a lot of symmetries, the main point of this Letter quite the opposite: we require \emph{less} symmetries than before. Formalism enables simulating distorted materials, but also reduces computational costs for certain simulations. Selected examples of usage are shown in Table~\ref{tab:usage}. For example, one cost-reduction area is surface science, where less atoms are needed to simulate thick surface slabs. We believe more application areas can be discovered, once the new concepts are mastered.

Now we leave the general discussion, and give one practical example of usage: we investigate twisted armchair graphene nanoribbons (AGNRs).\cite{son_PRL_06} We choose this example for the existing literature, but also for the possibility to illustrate operations beyond standard chiral symmetry.

\begin{figure}[tb]
\includegraphics[width=8cm]{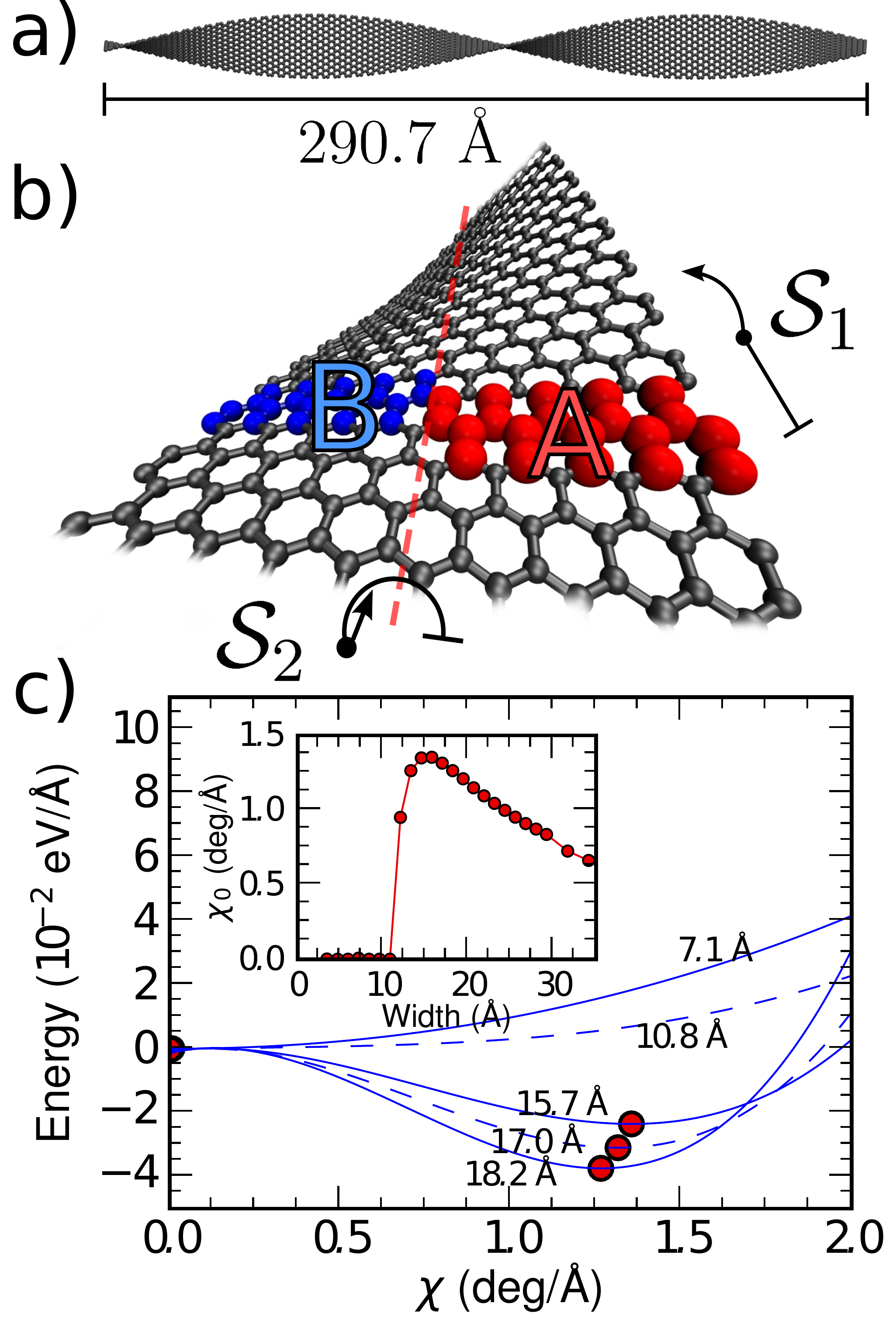}
\caption{(Color online) (a) Conventional unit cell of $20$-AGNR with a twist $\chi=1.2$~deg/\AA\ ($2720$ atoms). (b) Minimal unit cell (atoms A), with illustrations of $\mathcal{S}_1$ and $\mathcal{S}_2$. $\mathcal{S}_1$ is the chiral operation and $\mathcal{S}_2$ transforms A$\rightarrow$B. (c) Energies for AGNRs as a function of $\chi$ for selected ribbon widths. Minimum of the curve gives the spontaneous twist $\chi_0$. We used a $20\times 2$ ${\bm \kappa}$-point mesh. Inset: $\chi_0$ as a function of AGNR width.}
\label{fig:operations}
\end{figure}

Figure~\ref{fig:operations}a shows a piece of an infinitely long $20$-AGNR, with a twist $\chi=360$~deg$/290.7$~\AA$=1.2$~deg/\AA, within one conventional unit cell of $2720$ atoms. The minimal unit cell, enabled by the new formalism, in turn, has $20$ atoms (atoms A in Fig.~\ref{fig:operations}b), and is accompanied by two symmetry operations: $\mathcal{S}_1$ is $L=4.2$~\AA\ translation, followed by $L\cdot \chi=5.04$~degree rotation, and $\mathcal{S}_2$ is $L/2=2.1$~\AA\ translation, followed by $182.52$ degree rotation ($=180$~deg$+\chi \cdot L/2$). The whole system can be built from one unit cell by $\mathcal{S}_1^{m_1}\mathcal{S}_2^{m_2}$ with $m_1=0,\pm 1,\pm 2,\ldots$ and $m_2=0,1$. Note that $\mathcal{S}_2^2 = \mathcal{S}_1$ and hence $\kappa_2 = \kappa_1/2 + \pi m_2$, while $\kappa_1$ is freely sampled. (While this was our choice for the symmetry operations, another, and equally sufficient, choice would have been to use only $\mathcal{S}_2$ with $m_2=0,\pm 1,\pm 2,\ldots$.)\cite{symmetric-AGNR}


Table~\ref{tab:compare} shows the wall-clock times for selected simulations. The simulations with $\chi=0$ show that the new formalism has no computational overhead compared to translation. The simulations with $\chi=1.2$~deg/\AA\ show that finite twist affects simulation times with neither minimal nor chiral cells. The translational cell was too large for direct simulation, and the timing was estimated from the scaling law time$\sim$(system size)$^3$. Note that, by decreasing $\chi$, the translational cell size---along with its timing---could easily be grown indefinitely. Point here is that twists even smaller than $1.2$~deg/\AA\ will be required to investigate the relevant physics of a $20$-AGNR. Conventional quantum-mechanical simulation is practically impossible. We also remark that, given compatible ${\bm \kappa}$-point samplings, energies and forces from different types of cells are the same within floating-point precision.

\renewcommand{\arraystretch}{1.2}
\begin{table}[t]
\caption{Timings for $20$-AGNR with different twists. Time is the wall-clock time per molecular dynamics step, calculated with a standard present-day desktop computer. $^{\dagger)}$
Timing obtained from scaling law.}
\label{tab:compare}
\begin{tabular}{p{4.25cm}ccc}
\hline & \\[-10pt]\hline
\hspace{1cm}unit cell & atoms & $\chi$ (deg/\AA) & time (s) \\
\hline

minimal (like A in Fig.~\ref{fig:operations}b)&
$20$ &
$0$ &
$1.51$  \\

chiral (like A+B in Fig.~\ref{fig:operations}b)&
$40$ &
$0$ &
$2.75$ \\

translational (like chiral)&
$40$ &
$0$ &
$2.75$  \\

minimal (A in Fig.~\ref{fig:operations}b)&
$20$ &
$1.2$ &
$1.51$  \\

chiral (A+B in Fig.~\ref{fig:operations}b)&
$40$ &
$1.2$ &
$2.75$ \\

translational (Fig.~\ref{fig:operations}a)&
$2720$ &
$1.2$ &
$9.26\times 10^4$ $^{\dagger)}$ \\
\hline & \\[-10pt]\hline
\end{tabular}
\end{table}

Figure~\ref{fig:operations}c shows AGNRs' energies as a function of twist. For ribbons wider than $\sim 12$~\AA\ the energy is at minimum with non-zero $\chi=\chi_0$; ribbons twist spontaneously. This confirms earlier predictions by classical potentials (using thousands of atoms) \cite{bets_NR_09,bhuang_PRL_09} and finite element modeling\cite{shenoy_PRL_08}. The physical reason for twisting is the compressive edge stress that elongates edges with respect to ribbon's center.\cite{bhuang_PRL_09,jun_PRB_08,reddy_APL_09,bets_NR_09} The stress we get ($1.7$~eV/\AA) agrees well with the stress ($1.5$~eV/\AA) from previous density-functional calculations.\cite{bhuang_PRL_09} For wide ribbons we get scaling $\chi_0\sim 22$~deg/width, and the difference to classical scaling $\chi_0\sim 18$~deg/width of Ref.~\onlinecite{bets_NR_09} comes mainly from quantum mechanics: the edge stress resides not only at the edge, but extends more into ribbon's center---a feature hard to reproduce by classical potentials.\cite{bhuang_PRL_09} This is also why we have no spontaneous twist for narrow ribbons. Ribbons $\sim 11$~\AA\ wide have nearly zero torsion constant, and could be used in ultrasensitive  torsion balances. Ultimately, very wide ribbons should show bifurcation into flat ribbons with ripples at the edges, but we won't discuss that here.\cite{bets_NR_09,shenoy_PRL_08}

\begin{figure}
\includegraphics[width=7.5cm]{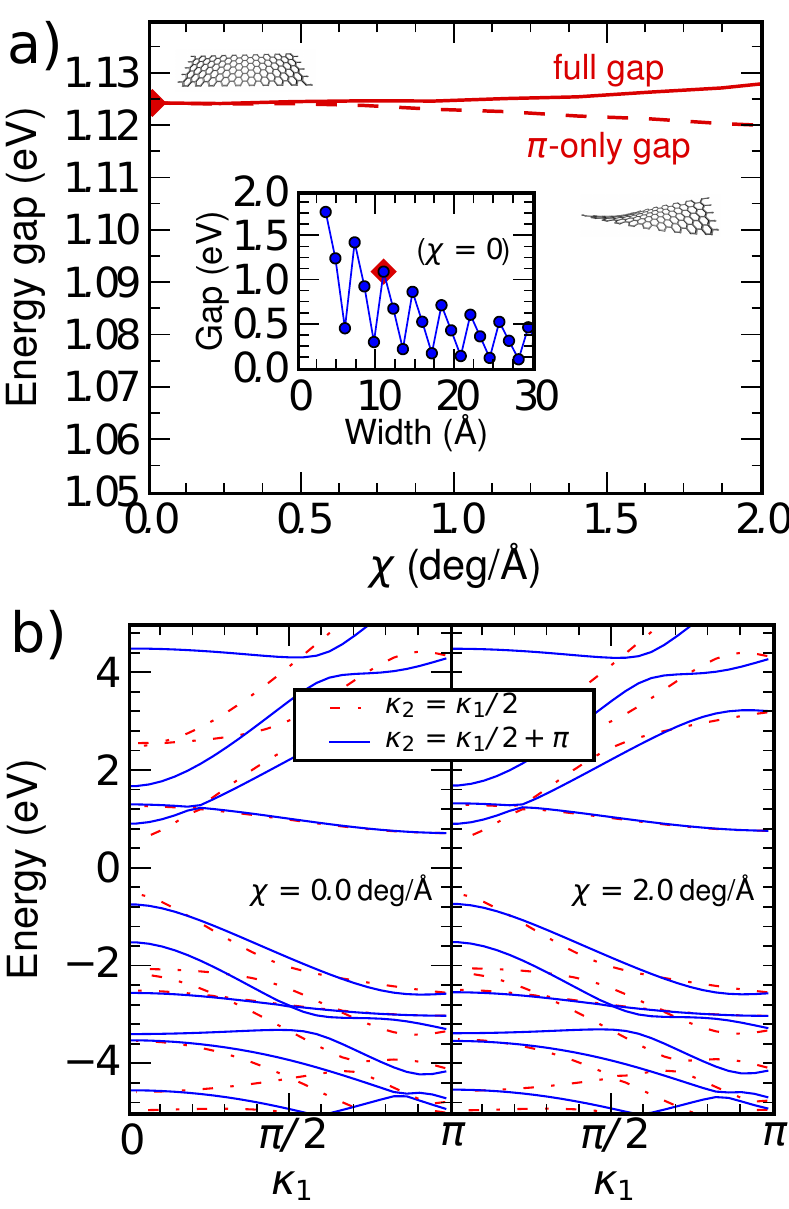}
\caption{(Color online) (a) Energy gap of a $10.8$~\AA\ wide $10$-AGNR as a function of $\chi$. The $\pi$-only gap is obtained by treating the unhybridized $2p_z$-electrons as $s$-electrons. Inset: energy gaps as a function of AGNR width for $\chi=0$. (b) Band structures for flat and twisted ribbons, plotted as a function of $\kappa_1$. The symmetry operation $\mathcal{S}_2$ brings another dimension, $\kappa_2$, to band structure plots. Calculation does not include spin.}
\label{fig:gaps}
\end{figure}

As argued in the abstract, the electronic properties ought to be investigated together with mechanical properties; this requires quantum mechanics. It is known that AGNRs have a gap due to the confinement of the finite width, and our gaps (inset in Fig.~\ref{fig:gaps}a) agree well with density-functional calculations of Ref.~\onlinecite{son_PRL_06}. But what happens to electronic structure when ribbons get twisted? Fig.~\ref{fig:gaps}a shows that twisting changes $10$-AGNR's gap very little---this is generic for all AGNRs. The gap from $\pi$-electrons alone shows further that $sp$-rehybridization is negligible. Even the band structures of flat and twisted ribbons (Fig.~\ref{fig:gaps}b) are nearly identical. This suggests that, contrary to CNTs\cite{yang_PRL_00}, the electronic properties of GNRs are remarkably robust against twisting.

To conclude, we hope to have illustrated how modest revision of Bloch's theorem enables versatile material distortions with quantum mechanics included, both numerically and analytically. However, excess emphasis on quantum mechanics causes undue discrimination of classical methods---the formalism works equally with classical force fields, finite element methods, or coarse-grained simulations, and equally when applied to, say, liquid-phase cell membranes, fluid flow through bent pipes, or electron transport.

We acknowledge the Academy of Finland for funding, H. H\"akkinen for support and the Finnish IT Center for Science (CSC) for computational resources.


\begin{thebibliography}{25}
\bibitem[{ABB()}]{ABBA-note}
\bibinfo{note}{The reflection is locally piecewise inside unit cells and works
  in a $\Gamma$-point calculation.}
\bibitem[{sym()}]{symmetric-AGNR}
\bibinfo{note}{With local basis the "unit cell" concept need not exist; we only
  have atom positions and $\mathcal{S}$'s (atoms can be spatially scattered,
  they need not be the closest images). Because we use local basis, ribbons
  where symmetry axis splits atoms need double the minimal cell.}
\end{thebibliography}

\end{document}